\documentclass[11pt]{article}

\usepackage{amsmath,amssymb}
\usepackage{array}
\usepackage{epsfig}

\usepackage{amsmath}
\usepackage{graphicx}
\usepackage{latexsym}

\newcount\figureno     \figureno=0
\newdimen\figdim       \figdim=70mm
\def\figureinc{%
   \global\advance\figureno by 1%
}
\def\figcaption#1#2#3{\hbox to #2{\hss{\vbox{\hsize=#2 \parindent=0pt
        {\bf Figure \number\figureno#3 :\ }#1}}\hss}
}

\begin{document}
\baselineskip 100pt

{\large
\parskip.2in
\newcommand{\be}{\begin{equation}}
\newcommand{\ee}{\end{equation}}
\newcommand{\ben}{\begin{equation*}}
\newcommand{\een}{\end{equation*}}
\newcommand{\br}{\bar}
\newcommand{\fr}{\frac}
\newcommand{\lm}{\lambda}
\newcommand{\ra}{\rightarrow}
\newcommand{\al}{\alpha}
\newcommand{\bt}{\beta}
\newcommand{\z}{\zeta}
\newcommand{\pa}{\partial}
\newcommand{\hs}{\hspace{5mm}}
\newcommand{\up}{\upsilon}
\newcommand{\dg}{\dagger}
\newcommand{\sdil}{\ensuremath{\rlap{\raisebox{.15ex}{$\mskip
6.5mu\scriptstyle+ $}}\subset}}
\newcommand{\sdir}{\ensuremath{\rlap{\raisebox{.15ex}{$\mskip
6.5mu\scriptstyle+ $}}\supset}}
\newcommand{\vphi}{\vec{\varphi}}
\newcommand{\ve}{\varepsilon}
\newcommand{\acc}{\\[3mm]}
\newcommand{\dl}{\delta}
\def\tablecap#1{\vskip 3mm \centerline{#1}\vskip 5mm}
\def\p#1{\partial_#1}
\newcommand{\pd}[2]{\frac{\partial #1}{\partial #2}}
\newcommand{\pdn}[3]{\frac{\partial #1^{#3}}{\partial #2^{#3}}}
\def\DP#1#2{D_{#1}\varphi^{#2}}
\def\dP#1#2{\partial_{#1}\varphi^{#2}}
\def\xh{\hat x}
\newcommand{\Ref}[1]{(\ref{#1})}

\def\mod#1{ \vert #1 \vert }
\def\chapter#1{\hbox{Introduction.}}
\def\Sin{\hbox{sin}}
\def\Cos{\hbox{cos}}
\def\Exp{\hbox{exp}}
\def\Ln{\hbox{ln}}
\def\Tan{\hbox{tan}}
\def\Cot{\hbox{cot}}
\def\Sinh{\hbox{sinh}}
\def\Cosh{\hbox{cosh}}
\def\Tanh{\hbox{tanh}}
\def\Asin{\hbox{asin}}
\def\Acos{\hbox{acos}}
\def\Atan{\hbox{atan}}
\def\Asinh{\hbox{asinh}}
\def\Acosh{\hbox{acosh}}
\def\Atanh{\hbox{atanh}}
\def\frac#1#2{{\textstyle{#1\over #2}}}

\newcommand{\ie}{{\it i.e.}}
\newcommand{\cmod}[1]{ \vert #1 \vert ^2 }
\newcommand{\cmodn}[2]{ \vert #1 \vert ^{#2} }
\newcommand{\nhat}{\mbox{\boldmath$\hat n$}}
\nopagebreak[3]
\bigskip

\title{ \bf Invariant solutions of the supersymmetric sinh--Gordon equation }
\vskip 1cm

\bigskip
\author{
A.~M. Grundland\thanks{email address: grundlan@crm.umontreal.ca}
\\
Centre de Recherches Math{\'e}matiques, Universit{\'e} de Montr{\'e}al,\\
C. P. 6128, Succ.\ Centre-ville, Montr{\'e}al, (QC) H3C 3J7,
Canada\\ Universit\'{e} du Qu\'{e}bec, Trois-Rivi\`{e}res, CP500 (QC) G9A 5H7, Canada \acc A. J. Hariton\thanks{email address: hariton@crm.umontreal.ca}
\\
Centre de Recherches Math{\'e}matiques, Universit{\'e} de Montr{\'e}al, \\
C. P. 6128, Succ.\ Centre-ville, Montr{\'e}al, (QC) H3C 3J7, Canada \acc L. \v{S}nobl\thanks{email address: Libor.Snobl@fjfi.cvut.cz}
\\
Faculty of Nuclear Sciences and Physical Engineering, \\
Czech Technical University in Prague, \\
Brehova 7, 115 19 Prague 1, Czech Republic} 
\date{}

\maketitle

\begin{abstract}

A systematic group-theoretical analysis of the supersymmetric sinh-Gordon equation is performed. A generalization of the method of prolongations is used to determine the Lie superalgebra of symmetries, and the method of symmetry reduction is applied in order to obtain invariant solutions of the supersymmetric sinh-Gordon equation. The results are compared with those previously found for the supersymmetric sine-Gordon equation. The presence of nonstandard invariants is discussed for the supersymmetric sinh-Gordon equation, as well as for the supersymmetric Korteweg-de Vries equation.

\end{abstract}




\newpage

\section{Introduction}

Recently, there has been much interest in the study of supersymmetric extensions of both classical and quantum mechanical models \cite{Grammaticos,Siddiq1,Siddiq2,Chaichian}. Various techniques have been used in order to obtain supersolitonic solutions, including the inverse scattering method, B\"{a}cklund transformations and their Riccati forms, Darboux transformations for the odd and even superfields, Lax formalism in superspace, and generalized versions of the symmetry reduction method \cite{Ablowitz1,Novikov,Lamb,Gardner,Zakharov,Ablowitz2,Lax}. Integrable models which were studied using these methods include (among others) the Korteweg-de Vries equation \cite{Mathieu,Labelle,HusAyaWin}, the sine-Gordon equation \cite{Grammaticos}, Liouville theory \cite{Chaichian}, the Schr\"{o}dinger equation \cite{Henkel} and sigma models \cite{HusZakr}. A number of solitonic and super multi-solitonic solutions were determined by a Crum-type transformation \cite{Matveev} and it was found \cite{Siddiq1,Siddiq2} that there exist infinitely many local conserved quantities. A connection was established between the super-Darboux transformations and super-B\"{a}cklund transformations, which allows one to construct $N$ super soliton solutions.

In a previous article \cite{SSG}, the Lie symmetry superalgebra of the supersymmetric sine-Gordon equation was determined by means of a generalization of the prolongation method and its subalgebras were classified. A number of invariant solutions of the model were found, including constant, algebraic, hyperbolic and doubly periodic solutions expressed in terms of elliptic functions. It was also found that some of the subalgebras had invariants which possess a non-standard structure in the sense that they do not admit symmetry reduction in the classical sense. 

The purpose of this paper is to perform a systematic study of the supersymmetric sinh-Gordon equation in order to determine its symmetry properties and invariant solutions. In order to do this, we proceed to employ the same group-theoretical techniques which we previously applied to the supersymmetric sine-Gordon equation \cite{SSG}. We also compare the Lie symmetry superalgebras and invariant solutions obtained for the supersymmetric sinh-Gordon equation to their counterparts which we obtained previously for the supersymmetric sine-Gordon equation. It will also be demonstrated that the Lie superalgebra of the supersymmetric Korteweg-de Vries equation also admits subalgebras with nonstandard invariants.

This paper is organized as follows. In section 2 we describe the supersymmetric version of the sinh-Gordon equation and the associated formalism. In section 3, we use a generalized form of the prolongation method to calculate the Lie superalgebra of symmetries of the supersymmetric sinh-Gordon equation. In section 4, we use the symmetry reduction method to calculate invariant solutions of the supersymmetric sinh-Gordon equation and discuss the fact that some subalgebras possess nonstandard invariants. We also discuss the nonstandard invariants for the case of the supersymmetric Korteweg-de Vries equation. Finally, in section 5 we present our conclusions, final remarks and possibilities for future research.

\section{Supersymmetric extension}

In order to supersymmetrize the sinh-Gordon equation
\begin{equation}
u_{xt}=\sinh{u},
\label{sshg1}
\end{equation}
we extend the space of independent variables $\{x,t\}$ to the superspace $\{x,t,\theta_1,\theta_2\}$, where $\theta_1$ and $\theta_2$ are independent odd (Grassmann) variables. We also replace the classical even (Grassmann) field $u(x,t)$ by the even (Grassmann) superfield
\begin{equation}
\Phi(x,t,\theta_1,\theta_2)=\frac{1}{2}u(x,t)+\theta_1\phi(x,t)+\theta_2\psi(x,t)+\theta_1\theta_2F(x,t),
\label{sshg2}
\end{equation}
where $\phi$ and $\psi$ are two new odd fields and $F$ is a new even field. We intend to construct the supersymmetric extension of (\ref{sshg1}) in such a way that it is invariant under the following two supersymmetry transformations:
\begin{equation}
x\rightarrow x-\underline{\eta}_1\theta_1,\quad \theta_1\rightarrow\theta_1+\underline{\eta}_1\qquad\mbox{ and }\qquad t\rightarrow t+\underline{\eta}_2\theta_2,\quad \theta_2\rightarrow\theta_2+\underline{\eta}_2,
\label{sshg3}
\end{equation}
where $\underline{\eta_1}$ and $\underline{\eta_2}$ are arbitrary constant odd parameters (where we use the convention that underlined constants represent odd parameters). The choice of signs in equation (\ref{sshg3}) is not arbitrary -- it is caused by the requirement that the supersymmetric version of the sinh--Gordon equation be written as an equation for real superfield, i.e. that the functions $u(x,t),\phi(x,t),\psi(x,t),F(x,t)$ take values in a real Grassmann ring and can therefore be physically interpreted as real bosonic and fermionic fields.

The two transformations (\ref{sshg3}) are generated by the infinitesimal supersymmetry operators
\begin{equation}
Q_x=\partial_{\theta_1}-\theta_1\partial_x\qquad\mbox{ and }\qquad Q_t=\partial_{\theta_2}+\theta_2\partial_t,
\label{sshg4}
\end{equation}
respectively. In order to make the generalized model invariant under the supersymmetry generators $Q_x$ and $Q_t$, we introduce the covariant derivatives
\begin{equation}
D_x=\partial_{\theta_1}+\theta_1\partial_x\qquad\mbox{ and }\qquad D_t=\partial_{\theta_2}-\theta_2\partial_t,
\label{sshg5}
\end{equation}
which possess the property that each derivative $D_i$ anticommutes with every supersymmetry operator $Q_j$. Also,
\begin{equation}
\{Q_x,Q_x\}=-2\partial_x,\quad \{Q_t,Q_t\}=2\partial_t,\quad \{Q_x,Q_t\}= \{D_x,D_t\}=0.
\label{sshg5A}
\end{equation}
Thus, if we write our supersymmetric equation in terms of the superfield $\Phi$ and its covariant derivatives of various orders, it will indeed be invariant under the transformations $Q_x$ and $Q_t$.
The superspace Lagrangian density of the supersymmetric model is
\begin{equation}
{\mathcal L}(\Phi)={1\over 2}D_x\Phi D_t\Phi+\cosh{\Phi},
\label{b6}
\end{equation}
and the corresponding Euler-Lagrange superfield equation (the supersymmetric sinh-Gordon equation) takes the form
\begin{equation}
D_xD_t\Phi=\sinh{\Phi}.
\label{sshg6}
\end{equation}
In terms of the partial derivatives with respect to the independent variables, this equation can be re--written as
\begin{equation}
-\theta_1\theta_2\Phi_{xt}+\theta_2\Phi_{t\theta_1}+\theta_1\Phi_{x\theta_2}-\Phi_{\theta_1\theta_2}-\sinh{\Phi}=0.
\label{sshg7}
\end{equation}

In this paper, we use the convention that for partial derivatives involving odd variables,
\begin{equation}
\partial_{\theta_i}(fg)=(\partial_{\theta_i}f)g+(-1)^{\mbox{deg}(f)}f(\partial_{\theta_i}g),
\label{convention}
\end{equation}
where
\begin{equation}
\mbox{deg}(f)=\begin{cases}0\mbox{ if }f\mbox{ is even}\\1\mbox{ if }f\mbox{ is odd}\end{cases}
\label{degree}
\end{equation}
and the notation
\begin{equation}
f_{\theta_1\theta_2} = \partial_{\theta_2} ( \partial_{\theta_1}f).  
\end{equation}

The even (super)numbers, variables, fields etc. are assumed to be elements of the even part $\Lambda_{even}$ of the underlying abstract real Grassmann ring $\Lambda=\wedge[\xi_1,\xi_2,\ldots]$; the odd (super)numbers, variables fields, etc. lie in its odd part $\Lambda_{odd}$. We shall assume throughout the paper that the function $u(x,t)$ in Eq. (\ref{sshg2}) (and $\alpha(\sigma)$ closely related to $u(x,t)$, used in Section \ref{invsolns}) has values in the invertible subset of $\Lambda_{even}$ plus $\{0 \}$, i.e. nonvanishing nilpotent values of $u(x,t)$ are ruled out. This technical assumption allows us to perform necessary simplifications in our calculations without splitting off of singular subcases.

\section{Symmetries of the supersymmetric sinh-Gordon equation}

We apply the generalized version of the method of prolongation of vector fields, as considered for Grassmann--valued systems of partial differential equations (see \cite{SSG}). That is, we postulate an even vector field of the form
\begin{equation}
\begin{split}
\mathbf{v}=&\xi(x,t,\theta_1,\theta_2,\Phi)\partial_x+\tau(x,t,\theta_1,\theta_2,\Phi)\partial_t+\rho(x,t,\theta_1,\theta_2,\Phi)\partial_{\theta_1}\\ &+\sigma(x,t,\theta_1,\theta_2,\Phi)\partial_{\theta_2}+\Lambda(x,t,\theta_1,\theta_2,\Phi)\partial_{\Phi},
\end{split}
\label{symmie3}
\end{equation}
where $\xi$, $\tau$ and $\Lambda$ are even functions, while $\rho$ and $\sigma$ are odd. We use the generalized total derivatives defined in \cite{SSG} in order to calculate the coefficients of the second prolongation of the vector field (\ref{symmie3}). In our case, the prolongation coefficients are exactly the same as those found for the supersymmetric sine-Gordon equation \cite{SSG}. We apply the second prolongation to the supersymmetric sinh-Gordon equation (\ref{sshg7}) in order to obtain the condition relating the various prolongation coefficients to each other.

Substituting the formulae for the prolongation coefficients into this condition and replacing each term $\Phi_{\theta_1\theta_2}$ in the resulting expression by the terms $-\theta_1\theta_2\Phi_{xt}+\theta_2\Phi_{t\theta_1}+\theta_1\Phi_{x\theta_2}-\sinh{\Phi}$, we obtain a series of determining equations for the functions $\xi$, $\tau$, $\rho$, $\sigma$ and $\Lambda$. The general solution of these determining equations is given by
\begin{equation}
\begin{split}
\xi(x,\theta_1)&=-2C_1x+C_2-\underline{D_1}\theta_1,\qquad
\tau(t,\theta_2)=2C_1t+C_3+\underline{D_2}\theta_2,\\
\rho(\theta_1)&=-C_1\theta_1+\underline{D_1},\qquad
\sigma(\theta_2)=C_1\theta_2+\underline{D_2},\qquad
\Lambda=0,
\end{split}
\label{symmie8}
\end{equation}
where $C_1$, $C_2$, $C_3$ are bosonic constants, while $\underline{D_1}$ and $\underline{D_2}$ are fermionic constants. Thus, we obtain that the superalgebra $\mathfrak{S}$ of symmetries of the supersymmetric sinh-Gordon equation (\ref{sshg7}) is the Poincar\'{e} superalgebra $P(1|1)$ generated by the following five infinitesimal vector fields:
\begin{equation}
\begin{split}
L=-2x\partial_x&+2t\partial_t-\theta_1\partial_{\theta_1}+\theta_2\partial_{\theta_2},\qquad P_x=\partial_x,\qquad P_t=\partial_t,\\ Q_x&=-\theta_1\partial_x+\partial_{\theta_1},\qquad Q_t=\theta_2\partial_t+\partial_{\theta_2}.
\end{split}
\label{sshg8}
\end{equation}
The generators $P_x$ and $P_t$ represent translations in space and time respectively, while $L$ generates a dilation in both even and odd independent variables. In addition, we recover the supersymmetry transformations $Q_x$ and $Q_t$ which we identified previously in (\ref{sshg4}). As we could expect by analogy with the non--supersymmetric case, no additional symmetries are obtained and this superalgebra is almost identical to the one which was found for the supersymmetric sine-Gordon equation -- it differs only by the sign of $\{ Q_t, Q_t \}$. This sign difference arises from the choice of supersymmetry generators and supercovariant derivatives in equations (\ref{sshg4}) and (\ref{sshg5}).
The commutation (anticommutation in the case of two odd operators) relations of the superalgebra $\mathfrak{S}$ of the supersymmetric sinh--Gordon equation  are given in Table 1.

The Lie superalgebra $\mathfrak{S}$ can be decomposed into the semi-direct sum
\begin{equation}
\mathfrak{S}=\{L\}\sdir\{P_x,P_t,Q_x,Q_t\}.
\label{sshg9}
\end{equation}
The classification of the one--dimensional subalgebras into conjugacy classes is similar to that found for the sine--Gordon equation (see \cite{SSG} for details) and is given as follows
\begin{equation}
\begin{split}
&\mathfrak{S}_1=\{L\},\\ &\mathfrak{S}_2=\{P_x\},\\ &\mathfrak{S}_3=\{P_t\},\\ &\mathfrak{S}_4=\{P_x+\varepsilon P_t\},\\ &\mathfrak{S}_5=\{\underline{\mu}Q_x\},\mbox{ 
where $\underline{\mu}$ and $\underline{\tilde\mu}=k \underline\nu$ represent the same conjugacy class }\\ & \hspace{2.7cm}\mbox{for any invertible even supernumber $k$ }\\ &\mathfrak{S}_6=\{P_x+\underline{\mu}Q_x\},\mbox{ 
where $\underline{\mu}$ and $\underline{\tilde\mu}=e^k \underline\nu$ represent the same conjugacy class}\\ & \hspace{3.6cm}\mbox{for any even supernumber $k$ }\\ &\mathfrak{S}_7=\{P_t+\underline{\mu}Q_x\},\mbox{ 
where $\underline\mu$ and $\underline{\tilde\nu}=e^k \underline{\nu}$ represent the same conjugacy class}\\ & \hspace{3.6cm}\mbox{for any even supernumber $k$ }\\ 
&\mathfrak{S}_8=\{P_x+\varepsilon P_t+\underline{\mu}Q_x\},
\\ &\mathfrak{S}_9=\{\underline{\nu}Q_t\},\mbox{ 
where $\underline{\mu}$ and $\underline{\tilde\mu}=k \underline\nu$ represent the same conjugacy class }\\ & \hspace{2.7cm}\mbox{for any invertible even supernumber $k$ }\\ 
&\mathfrak{S}_{10}=\{P_x+\underline{\nu}Q_t\},\mbox{ 
where $\underline\nu$ and $\underline{\tilde\nu}=e^k \underline\nu$ represent the same conjugacy class}\\ & \hspace{3.6cm}\mbox{for any even supernumber $k$ }\\ 
&\mathfrak{S}_{11}=\{P_t+\underline{\nu}Q_t\},\mbox{ 
where $\underline{\nu}$ and $\underline{\tilde\nu}=e^k \underline\nu$ represent the same conjugacy class}\\ & \hspace{3.6cm}\mbox{for any even supernumber $k$ }\\ 
&\mathfrak{S}_{12}=\{P_x+\varepsilon P_t+\underline{\nu}Q_t\},\\ &\mathfrak{S}_{13}=\{\underline{\mu}Q_x+\underline{\nu}Q_t\},\mbox{ 
where  $(\underline{\mu},\underline\nu)$ and $(\underline{\tilde\mu},\underline{\tilde\nu})=(e^k \underline{\mu},e^k\underline\nu)$ for any even}\\ & \hspace{3.6cm}\mbox{supernumber $k$ represent the same the conjugacy class }\\ &\mathfrak{S}_{14}=\{P_x+\underline{\mu}Q_x+\underline{\nu}Q_t\},\mbox{ 
where  $(\underline\mu,\underline\nu)$ and $(\underline{\tilde\mu}, \underline{\tilde\nu})=(e^k \underline\mu, e^{3 k} \underline\nu) $ represent the}\\ & \hspace{3.6cm}\mbox{same conjugacy class for any even supernumber $k$ }\\ 
&\mathfrak{S}_{15}=\{P_t+\underline{\mu}Q_x+\underline{\nu}Q_t\},\mbox{ 
where  $(\underline\mu,\underline\nu)$ and $(\underline{\tilde\mu}, \underline{\tilde\nu})=(e^{3k} \underline\mu, e^{k}\underline\nu) $ represent the}\\ & \hspace{3.6cm}\mbox{same conjugacy class for any even supernumber $k$ }\\ 
 &\mathfrak{S}_{16}=\{P_x+\varepsilon P_t+\underline{\mu}Q_x+\underline{\nu}Q_t\}.
\end{split}
\label{sshg10}
\end{equation}
These subalgebras allow us to determine invariant solutions of the supersymmetric sinh-Gordon equation (\ref{sshg7}).

\section{Invariant solutions}\label{invsolns}

We now proceed to apply a modified version of the symmetry reduction method to the supersymmetric sinh-Gordon equation (\ref{sshg7}) in order to obtain invariant solutions. Passing systematically through each subalgebra in the classification, we construct (where possible) a set of four functionally independent invariants. For the subalgebras $\mathfrak{S}_5$, $\mathfrak{S}_9$, $\mathfrak{S}_{13}$, $\mathfrak{S}_{14}$, $\mathfrak{S}_{15}$ and $\mathfrak{S}_{16}$, the invariants possess a non-standard structure, which will be discussed below, at the end of this section. They are the same as those found for the supersymmetric sine-Gordon equation \cite{SSG}. For the remaining subalgebras, the bosonic superfield $\Phi$ is written as a linear combination of the various invariants. That is, if the independent invariants are given by $\sigma$, $\tau_1$, $\tau_2$, where $\sigma$ is an even invariant while $\tau_1$ and $\tau_2$ are odd invariants, then the superfield $\Phi$ can be written in the form
\begin{equation}
\Phi={\mathcal A}(\sigma,\tau_1,\tau_2)=\alpha(\sigma)+\tau_1\eta(\sigma)+\tau_2\lambda(\sigma)+\tau_1\tau_2\beta(\sigma),
\label{sshg11}
\end{equation}
where $\alpha$ and $\beta$ are even-valued functions, while $\eta$ and $\lambda$ are odd-valued functions to be determined. When this decomposition is substituted into the supersymmetric sinh-Gordon equation, we obtain a reduced system of ordinary differential equations for the functions $\alpha$, $\eta$, $\lambda$ and $\beta$. In general, the term $\sinh{\mathcal A}$ can be expanded into the form
\begin{equation}
\sinh{\mathcal A}=(\sinh{\alpha})+\tau_1\eta(\cosh{\alpha})+\tau_2\lambda(\cosh{\alpha})+\tau_1\tau_2\left(\beta(\cosh{\alpha})-\eta\lambda(\sinh{\alpha})\right),
\label{sshg12}
\end{equation}
as identified by the series:
\begin{equation}
\sinh{\mathcal A}={\mathcal A}+{1\over 3!}{\mathcal A}^3+{1\over 5!}{\mathcal A}^5 + \ldots
\label{ginv3}
\end{equation}
We summarize our results as follows. In Table 2, we list the invariants of the respective one-dimensional subalgebras together with the form of their superfield solutions. In Table 3, we present the respective reduced systems of ordinary differential equations for $\alpha$, $\eta$, $\lambda$ and $\beta$.

For the sake of simplicity we unify the notation as follows: $\alpha$ and $\beta$ are even functions of their arguments, while $\eta$ and $\lambda$ are odd functions of their argument.

Subalgebras $\mathfrak{S}_5=\{\underline{\mu}Q_x\}$, $\mathfrak{S}_9=\{\underline{\nu}Q_t\}$, $\mathfrak{S}_{13}=\{\underline{\mu}Q_x+\underline{\nu}Q_t\}$, $\mathfrak{S}_{14}=\{P_x+\underline{\mu}Q_x+\underline{\nu}Q_t\}$, $\mathfrak{S}_{15}=\{P_t+\underline{\mu}Q_x+\underline{\nu}Q_t\}$, $\mathfrak{S}_{16}=\{P_x+\varepsilon P_t+\underline{\mu}Q_x+\underline{\nu}Q_t\}$ have invariants which possess non-standard structures and will be discussed at the end of this section.

For subalgebras $\mathfrak{S}_2=\{P_x\}$, $\mathfrak{S}_3=\{P_t\}$, $\mathfrak{S}_6=\{P_x+\underline{\mu}Q_x\}$, $\mathfrak{S}_7=\{P_t+\underline{\mu}Q_x\}$, $\mathfrak{S}_{10}=\{P_x+\underline{\nu}Q_t\}$ and $\mathfrak{S}_{11}=\{P_t+\underline{\nu}Q_t\}$ the only solution of the reduced equations is the null solution $\Phi=0$.

Subalgebra $\mathfrak{S}_1=\{L\}$ leads to the solution
\begin{equation}
\Phi=\alpha(\sigma)+t^{1/2}\theta_1\eta(\sigma)+t^{-1/2}\theta_2\lambda(\sigma)+\theta_1\theta_2\beta(\sigma),
\label{sol1}
\end{equation}
where the symmetry variable is $\sigma=xt$, the functions $\alpha$ and $\lambda$ satisfy the following ordinary differential equations
\begin{equation}
\begin{split}
& \alpha_{\sigma\sigma}+\sigma^{-1}\alpha_{\sigma}-\frac{1}{2}\sigma^{-1}\sinh{(2\alpha)}-C_0\sigma^{-3/2}\sinh{\alpha}=0,\\
& \lambda_{\sigma\sigma}+(\frac{1}{2}\sigma^{-1}-(\tanh{\alpha})\alpha_{\sigma})\lambda_{\sigma}-\sigma^{-1}\cosh^2{\alpha}\lambda=0,
\end{split}\label{sol1A}
\end{equation}
and $\eta$, $\beta$ are expressed as
\begin{equation}
\begin{split}
& \eta={1\over \cosh{\alpha}}\lambda_{\sigma},\\
& \beta=-\sinh{\alpha},
\end{split}
\end{equation}
subject to the condition that 
\begin{equation}
\lambda\eta=C_0\sigma^{-1/2},
\label{sol1B}
\end{equation}
where $C_0$ is a nilpotent even constant. This represents a nontrivial scaling--invariant solution, where the ordinary differential equation for $\alpha$ does not have the Painlev\'{e} property and its solution in closed form is unknown.

The reduction with respect to the subalgebra $\mathfrak{S}_4=\{P_x+\varepsilon P_t\}$ implies that
$$ (\eta \lambda)_\sigma=0 , $$
i.e. $\eta \lambda=C_0$ is an even nilpotent constant. The bosonic part of the equations of motion becomes 
\begin{equation}\label{boseqalpha}
 \varepsilon\alpha_{\sigma\sigma}+\sinh{\alpha} \cosh{\alpha}+C_0 \sinh{\alpha}=0.
\end{equation}
Firstly, we restrict ourselves to $C_0=0$. This choice allows us to find a solution of the equation for $\alpha(\sigma)$ in the implicit form (\ref{sol4A}). Consequently, we find the following solution of the supersymmetric sinh-Gordon equation (\ref{sshg7})
\begin{equation}
\Phi=\alpha(\sigma)+\theta_1\eta(\sigma)+\theta_2\lambda(\sigma)+\theta_1\theta_2\beta(\sigma),
\label{sol4}
\end{equation}
where the symmetry variable is $\sigma=x-\varepsilon t$. The function $\alpha$ is expressed in terms of the elliptic function $F$:
\begin{equation}
\begin{split}
& {\pm \sqrt{2\varepsilon\cosh^2{\alpha}-\varepsilon-4C_1\over 4C_1+\varepsilon}\,F\left(\cosh{\alpha},\sqrt{2\varepsilon\over 4C_1+\varepsilon}\right)\over \sqrt{2C_1-\varepsilon\cosh^2{\alpha}+\frac{1}{2}\varepsilon}}=\sigma+C_2.
\end{split}
\label{sol4A}
\end{equation}
The function $\lambda$ has the form $\lambda=\underline{K}f(\sigma)$, where $\underline{K}$ is an odd constant and $f$ is an even function which satisfies the equation
\begin{equation}
\underline{K}\left[f_{\sigma\sigma}-(\tanh{\alpha})\alpha_{\sigma}f_{\sigma}+\varepsilon(\cosh^2{\alpha})f\right]=0.
\label{sol4AA}
\end{equation}
The functions $\lambda$ and $\beta$ are defined as follows
\begin{equation}
\begin{split}
& \eta={\underline{K}\over \cosh{\alpha}}f_{\sigma},\\
& \beta=-\sinh{\alpha}.
\end{split}
\label{sol4B}
\end{equation}

This represents a travelling wave expressed in terms of elliptic functions. We observe that, by choosing $\underline{K}=0$, we can make $\Phi$ in equation (\ref{sol4}) into a purely bosonic nontrivial solution, i.e. $\eta=\lambda=0$.

When we set $\eta \lambda=C_0\neq 0$ in equation (\ref{boseqalpha}) we find a more complicated implicit solution $\alpha(\sigma)$
\begin{equation}
\int_{a_0}^{a=\alpha(\sigma)}{ \pm 2e^{a}\over\sqrt{-e^{4a}-4C_0e^{3a}+(4C_1-2)e^{2a}-4C_0e^{a}-1}} {\rm d}a - \sigma = 0,
\label{complicatedsolution1}
\end{equation}
where $C_1,a_0$ are integration constants. The integral in equation (\ref{complicatedsolution1}) can be converted via the substitution $y=e^a$ to the elliptic integral, giving an equation
\begin{equation}
\int_{y_0}^{y(\sigma)}{ \pm 2{\rm d}y\over\sqrt{-y^4-4C_0y^3+(4C_1-2)y^2-4C_0y-1}}-\sigma = 0
\label{complicatedsolution2}
\end{equation}
The general solution of equation (\ref{complicatedsolution2}) is well known when $C_0,C_1$ are ordinary real numbers (see e.g. \cite{Whittaker} p. 453). The solution $y$ can be expressed as a rational Weierstrass elliptic function
\begin{equation}
y-y_0=\frac{1}{4}f'(y_0)\{{\mathcal P}(\sigma,g_2,g_3)-\frac{1}{24}f''(y_0)\}^{-1}, 
\label{weierstrassp}
\end{equation}
where the invariants of the elliptic Weierstrass function are
\begin{equation}
g_2=\frac{4}{3}-4C_0^2+\frac{4}{3}C_1(C_1-1),\qquad g_3=\frac{4}{9}C_1-\frac{8}{27}+\frac{2}{3}C_0^2C_1-\frac{7}{3}C_0^2-\frac{8}{27}C_1^3+\frac{4}{9}C_1^2,
\label{weierstrassinvariants}
\end{equation}
and the function $f$ is defined by $f(y)=-y^4-4C_0y^3+(4C_1-2)y^2-4C_0y-1$.
Depending on the values of $C_0$ and $C_1$, this can lead to doubly periodic solutions expressed in terms of the Jacobi elliptic functions $sn(\xi,k)$, $cn(\xi,k)$ and $dn(\xi,k)$. Due to the presence of the nilpotent even constant $C_0$, the constants $g_2,g_3$ and consequently also the modulus $k$ must be considered in the whole ring of even supernumbers $\Lambda_{even}$ and cannot be restricted to be real or complex numbers only. The properties of such a generalization of elliptic functions are, as far as we know, not yet fully understood and an understanding going much further than the standard references, e.g. \cite{Byrd}, would be required for a full analysis and explicit construction of the solution of the reduced equations. Nevertheless, under the assumption that these functions can be consistently generalized to Grassmann ring--valued parameters, i.e. in $\Lambda_{even}$, we conjecture that the solution of equation (\ref{complicatedsolution2}) retains the form (\ref{weierstrassp}) even for $C_0$ nilpotent.

The odd fields $\lambda,\eta$ are then solutions of the following homogeneous coupled linear ordinary differential equations
$$\lambda_{\sigma}-\eta\cosh{\alpha}=0,\qquad \varepsilon\eta_{\sigma}+\lambda\cosh{\alpha}=0$$
constrained by the condition $\eta \lambda=C_0$.

When reducing with respect to the subalgebra $\mathfrak{S}_8=\{P_x+\varepsilon P_t+\underline{\mu}Q_x\}$, i.e. considering the equations
\begin{equation}\label{redeqS8}
 \begin{split}
& \varepsilon\alpha_{\sigma\sigma}+\underline{\mu}\eta_{\sigma}+\sinh{\alpha}\cosh{\alpha}+\eta\lambda\sinh{\alpha}=0,  \\
& \varepsilon\lambda_{\sigma}-\eta\cosh{\alpha}=0, \qquad \eta_{\sigma}+\underline{\mu}\alpha_{\sigma}+\lambda\cosh{\alpha}=0
 \end{split}
\end{equation}
we arrive at the constraint
\begin{equation}\label{S8restr}
(\eta \lambda)_\sigma=-(\alpha_\sigma) \underline{\mu} \lambda.  
\end{equation}
A general solution of the coupled set of reduced equations (\ref{redeqS8}) is not known.
If we assume that both $\eta$ and $\lambda$ are multiples of $\underline{\mu}$ then equation (\ref{S8restr}) holds trivially and the differential equation for $\alpha$ becomes again 
$$ \varepsilon\alpha_{\sigma\sigma}+\sinh{\alpha}\cosh{\alpha}=0. $$
Its general solution is therefore the same as in equation (\ref{sol4A}).
Consequently, we arrive at the solution of the supersymmetric sinh-Gordon equation (\ref{sshg7})
\begin{equation}
\Phi=\alpha(\sigma)+(\theta_1-\varepsilon\underline{\mu}t)\eta(\sigma)+\theta_2\lambda(\sigma)+(\theta_1-\varepsilon\underline{\mu}t)\theta_2\beta(\sigma),
\label{sol8}
\end{equation}
where the symmetry variable is $\sigma=\varepsilon x-t+\underline{\mu}t\theta_1$, the function $\alpha$ is defined by equation (\ref{sol4A}),
$\lambda=\underline{\mu}f(\sigma)$, where $f$ is an even function which satisfies the inhomogeneous linear ordinary differential equation
\begin{equation}
\underline{\mu}\left[f_{\sigma\sigma}-(\tanh{\alpha})\alpha_{\sigma}f_{\sigma}+\varepsilon (\cosh^2{\alpha})f+\varepsilon(\cosh{\alpha})\alpha_{\sigma}\right]=0,
\label{sol8AA}
\end{equation}
\begin{equation}
\begin{split}
& \eta={\varepsilon\underline{\mu}\over \cosh{\alpha}}f_{\sigma},\\
& \beta=-\sinh{\alpha}.
\end{split}
\label{sol8B}
\end{equation}
This represents a travelling simple wave involving $x$, $t$ modified by the odd variable $\theta_1$. In this case, a solution with $\eta=\lambda=0$ is not present.

The reduction with respect to the subalgebra $\mathfrak{S}_{12}=\{P_x+\varepsilon P_t+\underline{\nu}Q_t\}$ proceeds similarly to the case $\mathfrak{S}_8=\{P_x+\varepsilon P_t+\underline{\mu}Q_x\}$. Under similar assumptions on $\lambda,\eta$, i.e. both of them being a multiple of $\underline{\nu}$ we find the solution
\begin{equation}
\Phi=\alpha(\sigma)+\theta_1\eta(\sigma)+(\theta_2-\underline{\nu}x)\lambda(\sigma)+\theta_1(\theta_2-\underline{\nu}x)\beta(\sigma),
\label{sol12}
\end{equation}
where the symmetry variable is $\sigma=t-\varepsilon x-\underline{\nu}x\theta_2$, the function $\alpha$ is defined by equation (\ref{sol4A}),
$\eta=\underline{\nu}f(\sigma)$, where $f$ is an even function which satisfies the inhomogeneous linear ODE
\begin{equation}
\underline{\nu}\left[f_{\sigma\sigma}-(\tanh{\alpha})\alpha_{\sigma}f_{\sigma}+\varepsilon(\cosh^2{\alpha})f-\varepsilon(\cosh{\alpha})\alpha_{\sigma}\right]=0,
\label{sol12AA}
\end{equation}
\begin{equation}
\begin{split}
& \lambda={\underline{\nu}\over \cosh{\alpha}}f_{\sigma},\\
& \beta=-\sinh{\alpha}.
\end{split}
\label{sol12B}
\end{equation}
The solution represents a travelling simple wave involving $x$, $t$ modified by the odd variable $\theta_2$. We note that the solutions for $\mathfrak{S}_8$ and $\mathfrak{S}_{12}$ are very similar -- one can be obtained from the other upon simultaneous interchange of $x$ nd $t$, $\theta_1$ and $\theta_2$, $\eta$ and $\lambda$, $\mu$ and $\nu$ and changes of signs which can be deduced from the difference in $Q_x$ and $Q_t$.

The elliptic function $F$ in equation (\ref{sol4A}) possesses one real and one purely imaginary period provided that the modulus
\begin{equation}
k=\frac{2\varepsilon}{4 C_1 +\varepsilon}
\label{modulusan}
\end{equation}
is such that $0<k^2<1$. This implies that either $C>\frac{1}{4}$ or $C<-\frac{3}{4}$ when $\varepsilon=1$ and similarly  $C<-\frac{1}{4}$ or $C<-\frac{3}{4}$ when $\varepsilon=-1$.

To sum up our results up to now, for subalgebras $\mathfrak{S}_1,\mathfrak{S}_4,\mathfrak{S}_{8},\mathfrak{S}_{12}$ we have obtained consistent reduced systems of equations which we were able to solve case by case under some additional assumptions about the form of the solution (where the solution may be implicit or involve a solution of a known linear ordinary differential equation whose coefficients depend on previously found, i.e. in principle known, functions). The subalgebras $\mathfrak{S}_2,\mathfrak{S}_3, \mathfrak{S}_6, \mathfrak{S}_7, \mathfrak{S}_{10}$  and $\mathfrak{S}_{11}$ allow consistent systems of reduced equations but their solution in each case is the null solution $\Phi=0$.

Those subalgebras whose invariants possess a non--standard structure, i.e. $\mathfrak{S}_5=\{\underline{\mu}Q_x\}$, $\mathfrak{S}_9=\{\underline{\nu}Q_t\}$, $\mathfrak{S}_{13}=\{\underline{\mu}Q_x+\underline{\nu}Q_t\}$, $\mathfrak{S}_{14}=\{P_x+\underline{\mu}Q_x+\underline{\nu}Q_t\}$, $\mathfrak{S}_{15}=\{P_t+\underline{\mu}Q_x+\underline{\nu}Q_t\}$ and $\mathfrak{S}_{16}=\{P_x+\varepsilon P_t+\underline{\mu}Q_x+\underline{\nu}Q_t\}$ are the same as those found for the supersymmetric sine-Gordon equation \cite{SSG}. Such subalgebras are distinguished by the fact that each of them admits an invariant expressed in terms of an arbitrary function of the superspace variables, multiplied by an odd constant. Such invariants are nilpotent and this causes complications in the computation. This aspect can be illustrated by means of the following example. The subalgebra $\mathfrak{S}_5=\{\underline{\mu}Q_x\}$ generates the first of the two one--parameter group transformations described in equation (\ref{sshg3}). Its invariants are $t$, $\theta_2$, $\Phi$ and any quantity of the form
\begin{equation}
\tau=\underline{\mu}f\left(x,t,\theta_1,\theta_2,\Phi\right),
\label{nonstandard1}
\end{equation}
where $f$ is an arbitrary function which can be either bosonic or fermionic. It is an open question as to whether or not a substitution of these invariants into the supersymmetric sinh-Gordon equation (\ref{sshg7}) can lead to a reduced system of equations expressible in terms of the invariants. This is clearly not possible for every function $f$. For example, in the case where $\tau=\underline{\mu}x\theta_1$, the system (\ref{sshg7}) transforms into the equation
\begin{equation}
\underline{\mu}x\theta_2{\mathcal A}_{t\tau}+\underline{\mu}x{\mathcal A}_{\tau\theta_2}+\sinh{\mathcal A}=0,
\label{nonstandard2}
\end{equation}
for the field
\begin{equation}
\Phi={\mathcal A}\left(t,\tau,\theta_2\right).
\label{nonstandard3}
\end{equation}
The presence of the variable $x$ in equation (\ref{nonstandard2}) demonstrates that we do not obtain a reduced equation expressible in terms of the invariants.
On the other hand, if we would like to perform the reduction with respect to the vector field $Q_x$ (i.e. without $\underline{\mu}$) we immediately find that it is not a subalgebra and  we have to reduce with respect to the two--dimensional subalgebra $\{Q_x,P_x\}$. That leads to $\Phi(t,\theta_2)$ and substituting into equation (\ref{sshg7}) we find the reduction
\begin{displaymath}\sinh{\Phi}=0,\end{displaymath}
which allows again only the null solution
\begin{equation}
 \Phi = 0.
\end{equation}

These non-standard invariants arise from the fact that, in the case where we allow both even and odd variables, it is not always possible to find a coordinate transformation which rectifies the vector fields.

It should be noted that nonstandard invariants exist also in the case of the $N=2$ supersymmetric Korteweg-de Vries equation \cite{HusAyaWin}
\begin{equation}
\begin{split}
&A_t+A_{xxx}-3a\theta_1\theta_2A_xA_{xx}-(a+2)\theta_1AA_{xx\theta_2}-(a+2)\left(\theta_1\theta_2AA_{xxx}-\theta_2AA_{xx\theta_1}\right)\\ &+(2a+1)\theta_2A_xA_{x\theta_1}+(a+2)\left(A_xA_{\theta_1\theta_2}+AA_{x\theta_1\theta_2}\right)-(2a+1)\theta_1A_xA_{x\theta_2}\\ &-(a-1)\left(\theta_1A_{\theta_2}A_{xx}-\theta_2A_{\theta_1}A_{xx}+A_{\theta_1}A_{x\theta_2}-A_{\theta_2}A_{x\theta_1}\right)-3aA^2A_x=0,
\end{split}
\label{SKdV}
\end{equation}
where $A(x,t,\theta_1,\theta_2)=u(x,t)+\theta_1\rho^1(x,t)+\theta_2\rho^2(x,t)+\theta_1\theta_2 v(x,t)$ is a bosonic superfield.
Here, the Lie symmetry superalgebra $\mathfrak{g}$ of the equation (\ref{SKdV}) is spanned by the generators \cite{HusAyaWin}
\begin{equation}
\begin{split}
\mathcal{C}_1=\partial_x,\qquad \mathcal{C}_2&=\partial_t,\qquad \mathcal{C}_3=x\partial_x+3t\partial_t+\frac{1}{2}\theta_1\partial_{\theta_1}+\frac{1}{2}\theta_2\partial_{\theta_2}-A\partial_A,\\ \mathcal{A}_1&=\theta_1\partial_x-\partial_{\theta_1},\qquad \mathcal{A}_2=\theta_2\partial_x-\partial_{\theta_2}.
\end{split}
\label{kdvsuperalg}
\end{equation}
There exist subalgebras of $\mathfrak{g}$ for which the invariants possess a nonstandard structure. For example, if we take the subalgebra $\underline{\mu}\mathcal{A}_1=\{\underline{\mu}\theta_1\partial_x-\underline{\mu}\partial_{\theta_1}\}$, the 
invariants are $t$, $\theta_2$, $\Phi$ and any quantity of the form
\begin{equation}
\tau=\underline{\mu}f\left(x,t,\theta_1,\theta_2,\Phi\right),
\label{nonstandard4}
\end{equation}
where $f$ is an arbitrary function which can be either bosonic or fermionic. Other examples include the subalgebra $\underline{\mu}\mathcal{A}_1+\underline{\nu}\mathcal{A}_2=\{(\underline{\mu}\theta_1+\underline{\nu}\theta_2)\partial_x-\underline{\mu}\partial_{\theta_1}-\underline{\nu}\partial_{\theta_2}\}$, for which the nonstandard invariant is $\underline{\mu}\underline{\nu}f\left(x,t,\theta_1,\theta_2,\Phi\right)$ and the subalgebra $\mathcal{C}_1-\underline{\mu}\mathcal{A}_1-\underline{\nu}\mathcal{A}_2=\{(1-\underline{\mu}\theta_1-\underline{\nu}\theta_2)\partial_x+\underline{\mu}\partial_{\theta_1}+\underline{\nu}\partial_{\theta_2}\}$, for which the nonstandard invariant is $\underline{\mu}\underline{\nu}f\left(t,\theta_1,\theta_2,\Phi\right)$.

\section{Final Remarks}

We have determined the Lie algebra of symmetries of the supersymmetric sinh--Gordon model and found that it is very similar to that of the supersymmetric sine--Gordon equation which we had previously determined.  Through the use of the symmetry reduction method we have constructed several exact analytic solutions of this model, including doubly periodic solutions in terms of Jacobi elliptic functions. There were fewer classes of nonvanishing invariant solutions for the supersymmetric sinh--Gordon than for its supersymmetric sine-Gordon counterpart. This is due to the fact that, in contrast to trigonometric functions (such as sin and cos) hyperbolic functions have very few roots. The solutions of the supersymmetric sinh--Gordon equation can be of use in determining solutions of the super-Korteweg-de Vries equations due to the link which exists between the two supersymmetric models \cite{Liu3}. It was found that both the supersymmetric sinh-Gordon equation and the supersymmetric Korteweg--de Vries equation admit nonstandard invariants. One open problem is to determine if all integrable supersymmetric systems possess nonstandard invariants in this way. Also, could we apply the group--theoretical methods used in this paper to other integrable equation of mathematical physics? Such equations would include, among others, the supersymmetric Schr\"{o}dinger equation (motivated by supersymmetric quantum mechanics \cite{Henkel,Crombrugghe}) and the supersymmetric Sawada-Kotera equation \cite{Tian}. These will be the subject of future investigations.

\subsection*{Acknowledgements}

The research of A.~M. Grundland and A. J. Hariton was supported by research grants from NSERC of Canada. The research of L. \v{S}nobl was supported by the research plan MSM6840770039 of the Ministry of Education of the Czech Republic.
A.~M. Grundland also thanks P. Exner and I. Jex for hospitality and the Doppler Institute project LC06002 of the Ministry of Education of the Czech Republic for support during his visit to the Czech Technical University in Prague.

{}

\begin{table}[htbp]
  \begin{center}
\caption{Supercommutation table for the Lie superalgebra $\mathfrak{S}$ spanned by the
  vector fields (\ref{sshg8})}
\vspace{5mm}
\setlength{\extrarowheight}{4pt}
\begin{tabular}{|c||c|c|c|c|c|}\hline
& $\mathbf{L}$ & $\mathbf{P_x}$ & $\mathbf{P_t}$ & $\mathbf{Q_x}$ & $\mathbf{Q_t}$\\[0.5ex]\hline\hline
$\mathbf{L}$ & $0$ & $2P_x$ & $-2P_t$ & $Q_x$ & $-Q_t$ \\\hline
$\mathbf{P_x}$ & $-2P_x$ & $0$ & $0$ & $0$ & $0$ \\\hline
$\mathbf{P_t}$ & $2P_t$ & $0$ & $0$ & $0$ & $0$ \\\hline
$\mathbf{Q_x}$ & $-Q_x$ & $0$ & $0$ & $-2P_x$ & $0$ \\\hline
$\mathbf{Q_t}$ & $Q_t$ & $0$ & $0$ & $0$ & $2P_t$ \\\hline
\end{tabular}
  \end{center}
\end{table}

\begin{table}[htbp]
  \begin{center}
\caption{Invariants and change of variable for subalgebras of the Lie superalgebra $\mathfrak{S}$ spanned by the
  vector fields (\ref{sshg8})}
\vspace{3mm}
\setlength{\extrarowheight}{4pt}
\begin{tabular}{|c|c|c|}\hline
Subalgebra & Invariants & Superfield\\[0.5ex]\hline\hline
$\mathfrak{S}_1=\{L\}$ & $\sigma=xt$, $\tau_1=t^{1/2}\theta_1$, & $\Phi={\mathcal A}\left(\sigma,\tau_1,\tau_2\right)=\alpha(\sigma)+\tau_1\eta(\sigma)+\tau_2\lambda(\sigma)+\tau_1\tau_2\beta(\sigma)$ \\
 & $\tau_2=t^{-1/2}\theta_2$, $\Phi$ &\\\hline
$\mathfrak{S}_2=\{P_x\}$ & $t$, $\theta_1$, $\theta_2$, $\Phi$  & $\Phi={\mathcal A}\left(t,\theta_1,\theta_2\right)=\alpha(t)+\theta_1\eta(t)+\theta_2\lambda(t)+\theta_1\theta_2\beta(t)$ \\\hline
$\mathfrak{S}_3=\{P_t\}$ & $x$, $\theta_1$, $\theta_2$, $\Phi$  & $\Phi={\mathcal A}\left(x,\theta_1,\theta_2\right)=\alpha(x)+\theta_1\eta(x)+\theta_2\lambda(x)+\theta_1\theta_2\beta(x)$ \\\hline
$\mathfrak{S}_4=\{P_x+\varepsilon P_t\}$ & $\sigma=x-\varepsilon t$, $\theta_1$, $\theta_2$, $\Phi$  & $\Phi={\mathcal A}\left(\sigma,\theta_1,\theta_2\right)=\alpha(\sigma)+\theta_1\eta(\sigma)+\theta_2\lambda(\sigma)+\theta_1\theta_2\beta(\sigma)$ \\\hline
$\mathfrak{S}_6=\{P_x+\underline{\mu}Q_x\}$ & $t$, $\tau_1=\theta_1-\underline{\mu}x$, $\theta_2$, $\Phi$  & $\Phi={\mathcal A}\left(t,\tau_1,\theta_2\right)=\alpha(t)+\tau_1\eta(t)+\theta_2\lambda(t)+\tau_1\theta_2\beta(t)$ \\\hline
$\mathfrak{S}_7=\{P_t+\underline{\mu}Q_x\}$ & $\sigma=x+\underline{\mu}\theta_1t$, & $\Phi={\mathcal A}\left(\sigma,\tau_1,\theta_2\right)=\alpha(\sigma)+\tau_1\eta(\sigma)+\theta_2\lambda(\sigma)+\tau_1\theta_2\beta(\sigma)$ \\
 & $\tau_1=\theta_1-\underline{\mu}t$, $\theta_2$, $\Phi$ & \\\hline
$\mathfrak{S}_8=\{P_x+\varepsilon P_t+\underline{\mu}Q_x\}$ & $\sigma=\varepsilon x-t+\underline{\mu}t\theta_1$, & $\Phi={\mathcal A}\left(\sigma,\tau_1,\theta_2\right)=\alpha(\sigma)+\tau_1\eta(\sigma)+\theta_2\lambda(\sigma)+\tau_1\theta_2\beta(\sigma)$ \\
& $\tau_1=\theta_1-\varepsilon\underline{\mu}t$, $\theta_2$, $\Phi$ & \\\hline
$\mathfrak{S}_{10}=\{P_x+\underline{\nu}Q_t\}$ & $\sigma=t-\underline{\nu}\theta_2x$, & $\Phi={\mathcal A}\left(\sigma,\theta_1,\tau_2\right)=\alpha(\sigma)+\theta_1\eta(\sigma)+\tau_2\lambda(\sigma)+\theta_1\tau_2\beta(\sigma)$ \\
 & $\theta_1$, $\tau_2=\theta_2-\underline{\nu}x$, $\Phi$ & \\\hline
$\mathfrak{S}_{11}=\{P_t+\underline{\nu}Q_t\}$ & $x$, $\theta_1$, $\tau_2=\theta_2-\underline{\nu}t$, $\Phi$  & $\Phi={\mathcal A}\left(x,\theta_1,\tau_2\right)=\alpha(x)+\theta_1\eta(x)+\tau_2\lambda(x)+\theta_1\tau_2\beta(x)$ \\\hline
$\mathfrak{S}_{12}=\{P_x+\varepsilon P_t+\underline{\nu}Q_t\}$ & $\sigma=t-\varepsilon x-\underline{\nu}x\theta_2$, & $\Phi={\mathcal A}\left(\sigma,\theta_1,\tau_2\right)=\alpha(\sigma)+\theta_1\eta(\sigma)+\tau_2\lambda(\sigma)+\theta_1\tau_2\beta(\sigma)$ \\
 & $\theta_1$, $\tau_2=\theta_2-\underline{\nu}x$, $\Phi$ &\\\hline
\end{tabular}
  \end{center}
\end{table}

\begin{table}[htbp]
  \begin{center}
\caption{Reduced Equations obtained for subalgebras of the Lie superalgebra $\mathfrak{S}$ spanned by the
  vector fields (\ref{sshg8})}
\vspace{3mm}
\setlength{\extrarowheight}{4pt}
\begin{tabular}{|c|c|}\hline
Subalgebra & Reduced Equations \\[0.5ex]\hline\hline
$\mathfrak{S}_1=\{L\}$ & $\beta+\sinh{\alpha}=0$,\qquad $\lambda_{\sigma}-\eta\cosh{\alpha}=0$,\\ & $\sigma\eta_{\sigma}+\frac{1}{2}\eta-\lambda\cosh{\alpha}=0$,\qquad $\alpha_{\sigma}+\sigma\alpha_{\sigma\sigma}+\beta\cosh{\alpha}-\eta\lambda\sinh{\alpha}=0$ \\\hline
$\mathfrak{S}_2=\{P_x\}$ & $\beta+\sinh{\alpha}=0$,\qquad $\eta\cosh{\alpha}=0$,\\ & $\eta_t-\lambda\cosh{\alpha}=0$,\qquad $\beta\cosh{\alpha}-\eta\lambda\sinh{\alpha}=0$ \\\hline
$\mathfrak{S}_3=\{P_t\}$ & $\beta+\sinh{\alpha}=0$,\qquad $\lambda_x-\eta\cosh{\alpha}=0$,\\ & $\lambda\cosh{\alpha}=0$,\qquad $\beta\cosh{\alpha}-\eta\lambda\sinh{\alpha}=0$ \\\hline
$\mathfrak{S}_4=\{P_x+\varepsilon P_t\}$ & $\beta+\sinh{\alpha}=0$,\qquad $\lambda_{\sigma}-\eta\cosh{\alpha}=0$,\\ & $\varepsilon\eta_{\sigma}+\lambda\cosh{\alpha}=0$,\qquad $\varepsilon\alpha_{\sigma\sigma}-\beta\cosh{\alpha}+\eta\lambda\sinh{\alpha}=0$ \\\hline
$\mathfrak{S}_6=\{P_x+\underline{\mu}Q_x\}$ & $\beta+\sinh{\alpha}=0$,\qquad $\underline{\mu}\beta-\eta\cosh{\alpha}=0$,\\ & $\eta_t-\lambda\cosh{\alpha}=0$,\qquad $\underline{\mu}\eta_t-\beta\cosh{\alpha}+\eta\lambda\sinh{\alpha}=0$ \\\hline
$\mathfrak{S}_7=\{P_t+\underline{\mu}Q_x\}$ & $\beta+\sinh{\alpha}=0$,\qquad $\lambda_{\sigma}-\eta\cosh{\alpha}=0$,\\ & $\underline{\mu}\alpha_{\sigma}+\lambda\cosh{\alpha}=0$,\qquad $\underline{\mu}\eta_{\sigma}-\beta\cosh{\alpha}+\eta\lambda\sinh{\alpha}=0$ \\\hline
$\mathfrak{S}_8=\{P_x+\varepsilon P_t+\underline{\mu}Q_x\}$ & $\beta+\sinh{\alpha}=0$,\qquad $\varepsilon\lambda_{\sigma}-\eta\cosh{\alpha}=0$,\\ & $\eta_{\sigma}+\underline{\mu}\alpha_{\sigma}+\lambda\cosh{\alpha}=0$,\qquad $\varepsilon\alpha_{\sigma\sigma}+\underline{\mu}\eta_{\sigma}-\beta\cosh{\alpha}+\eta\lambda\sinh{\alpha}=0$ \\\hline
$\mathfrak{S}_{10}=\{P_x+\underline{\nu}Q_t\}$ & $\beta+\sinh{\alpha}=0$,\qquad $\underline{\nu}\alpha_{\sigma}-\eta\cosh{\alpha}=0$,\\ & $\eta_{\sigma}-\lambda\cosh{\alpha}=0$,\qquad $\underline{\nu}\lambda_{\sigma}-\beta\cosh{\alpha}+\eta\lambda\sinh{\alpha}=0$ \\\hline
$\mathfrak{S}_{11}=\{P_t+\underline{\nu}Q_t\}$ & $\beta+\sinh{\alpha}=0$,\qquad $\lambda_x-\eta\cosh{\alpha}=0$,\\ & $\underline{\nu}\beta+\lambda\cosh{\alpha}=0$,\qquad $\underline{\nu}\lambda_x-\beta\cosh{\alpha}+\eta\lambda\sinh{\alpha}=0$ \\\hline
$\mathfrak{S}_{12}=\{P_x+\varepsilon P_t+\underline{\nu}Q_t\}$ & $\beta+\sinh{\alpha}=0$,\qquad $\underline{\nu}\alpha_{\sigma}-\varepsilon\lambda_{\sigma}-\eta\cosh{\alpha}=0$,\\ & $\eta_{\sigma}-\lambda\cosh{\alpha}=0$,\qquad $\varepsilon\alpha_{\sigma\sigma}+\underline{\nu}\lambda_{\sigma}-\beta\cosh{\alpha}+\eta\lambda\sinh{\alpha}=0$ \\\hline
\end{tabular}
  \end{center}
\end{table}

\label{lastpage}
\end{document}